\documentclass[aip,jcp,reprint]{revtex4-1} 
\usepackage{graphicx}
\usepackage{enumitem}

\usepackage{mathtools}
\usepackage{graphicx}   
\usepackage{color}
\usepackage[dvipsnames]{xcolor}
\usepackage{ulem}
\usepackage{caption}
\usepackage{subcaption}
\usepackage{comment}
\usepackage{bm}

\usepackage{natbib}

\usepackage{array}
\usepackage{booktabs}
\usepackage{multirow}

\usepackage{amsmath}

\usepackage{booktabs}

\DeclareMathOperator\erf{erf}
\DeclareMathOperator\erfc{erfc}

\begin{document}
\title{Quantifying the relevance of long-range forces for crystal nucleation in water}

 \author{Renjie Zhao}
 \affiliation{Chemical Physics Program and Institute for Physical Science and Technology, University of Maryland, College Park 20742, USA.}

\author{Ziyue Zou}
 \affiliation{Department of Chemistry and Biochemistry, University of Maryland, College Park 20742, USA.}

 \author{John D. Weeks\footnote{Corresponding author.}}
 \email{jdw@umd.edu}

\affiliation{Institute for Physical Science and Technology and Department of Chemistry and Biochemistry, University of Maryland, College Park 20742, USA.}

 \author{Pratyush Tiwary\footnote{Corresponding author.}}
 
 \email{ptiwary@umd.edu}
 \affiliation{Institute for Physical Science and Technology and Department of Chemistry and Biochemistry, University of Maryland, College Park 20742, USA.}

\begin{abstract}
Understanding nucleation from aqueous solutions is of fundamental importance in a multitude of fields, ranging from materials science to biophysics. The complex solvent-mediated interactions in aqueous solutions hamper the development of a simple physical picture elucidating the roles of different interactions in nucleation processes. In this work we make use of three complementary techniques to disentangle the role played by short and long-range interactions in solvent mediated nucleation. Specifically, the first approach we utilize is the local molecular field (LMF) theory to renormalize long-range Coulomb electrostatics. Secondly, we use well-tempered metadynamics to speed up rare events governed by short-range interactions. Thirdly, deep learning-based State Predictive Information Bottleneck approach is employed in analyzing the reaction coordinate of the nucleation processes obtained from LMF treatment coupled with well-tempered metadynamics.  We find that the two-step nucleation mechanism can largely be captured by the short-range interactions, while the long-range interactions further contribute to the stability of the primary crystal state at ambient conditions. Furthermore, by analyzing the reaction coordinate obtained from combined LMF-metadynamics treatment, we discern the fluctuations on different time scales, highlighting the need for long-range interactions when accounting for metastability.
\end{abstract}

	\maketitle
\section*{Introduction}
Nucleation from aqueous solutions has been extensively investigated in recent years due to its theoretical and practical importance. In comparison to nucleation from the melt, the presence of solvent molecules substantially enhances the complexity of the configurational phase space, introduces non-trivial finite-size effects \cite{salvalaglio2015molecular,karmakar2019molecular}, and requires more sophisticated methods in constructing reaction coordinates\cite{zou2023driving} (RCs) that can encapsulate the governing mechanism. The energy barriers of crystalline nucleation are typically significantly higher than the thermal fluctuations in the systems of interest. Consequently, nucleation is considered as a microscopically rare event which transpires over timescales that far surpasses the computational limits of molecular dynamics (MD) simulations. As the association of solute molecules involves breaking and reforming of solute-solvent and solvent-solvent hydrogen bonds, aqueous solutions present particular difficulty in overcoming nucleation barriers. In general, these short-range interactions play a major role in controlling the dynamics of such systems. On this account, advanced sampling methods that efficiently enhance the fluctuations arising from such short-range interactions have become prevalent tools in the study of nucleation processes and other rare events more generally.\cite{salvalaglio2015molecular,karmakar2019molecular,finney2022multiple,zou2023driving,anwar2011uncovering,tribello2009molecular,zahn2004atomistic,giberti2015insight,giberti2013transient,knott2012homogeneous,zou2021toward} A very active area of research in this context involves the development of suitable descriptors for approximating the RC, that can be biased in enhanced sampling approaches such as well-tempered metadynamics\cite{barducci2008well} (WTmetaD). For instance, these include approximations to thermodynamic descriptors,\cite{mendels2018searching,piaggi2017enhancing,piaggi2018predicting} and more recently, Artificial Intelligence (AI) based approaches.\cite{zou2021toward,zou2023driving} The aforementioned approaches can describe and enhance the short-range intermolecular interactions accounting for the local hydrogen bond network and local solvation structure in aqueous solutions. 

However, when it comes to nucleation in nonuniform charged and polar liquids, the long-range Coulomb interactions also play a different and pronounced role. Long wavelength constraints like neutrality and dielectric screening generally arise from the long-range tails of Coulomb interactions,\cite{stillinger1968general} independent of most details of the short-range structure. By exploiting the separation of Coulomb interactions into strong short-range and uniformly slowly varying long-range components, local molecular field (LMF) theory\cite{rodgers2008local} focuses on the very different roles the short and long-range components of intermolecular interactions in determining relevant structural,\cite{weeks2002connecting,chen2004connecting,rodgers2008interplay,rodgers2011efficient} thermodynamic,\cite{rodgers2009accurate,remsing2016long,gao2020short} and dynamical properties\cite{baker2020local,zhao2020response,wang2022influence} in nonuniform fluids. More specifically, based on the controlled approximations used in LMF theory, the short solvent (SS) model\cite{gao2020short,gao2023local} was developed to renormalize effective long-range interactions between intermolecular solute sites, while retaining only short-range interactions for other atomic site pairs. This approach significantly reduces the computational complexity for aqueous systems. The SS model has been applied to the association of ionic and hydrophobic solutes with great success.\cite{gao2020short,gao2023local} However, the investigation of rare events in more complicated systems, such as nucleation of molecular systems in water, poses additional challenges in discerning solvent-mediated effects, particularly when assessing the collective system response to long-range interactions over large time scales.

In order to achieve an efficient and physically insightful treatment for both short and long-range interactions involved in nucleation in solvated media, we propose a modeling approach using the LMF-based SS model for aqueous solutions of urea, coupled with WTmetaD to enhance the sampling of nucleation events. Urea is a well-known protein denaturant at high concentrations. Recently urea nucleation in water has been studied with advanced sampling methods, emphasizing the nucleation mechanism\cite{salvalaglio2015molecular} and the presence of competing polymorphs.\cite{zou2023driving} 

In this work, we aim to analyze the transitions between the liquid state and the experimental stable crystal structure of urea under ambient conditions, which is taken as the testing ground for the SS model’s ability to describe the nucleation processes in water and predict the corresponding free energy difference. The combination of LMF theory and WTmetaD allows us to disentangle the roles of short and long-range interactions in the collective response of the system to solvent-mediated interactions in the course of crystal nucleation and dissolution. With the observed multiple nucleation events in all-atom resolution, we then dive deep into the nucleation mechanism. Classical nucleation theory\cite{gibbs1876equilibrium} (CNT) is considered inadequate for describing nucleation phenomena from solutions due to its reliance on the assumption that the nucleation process occurs primarily along  the size of the largest crystalline cluster.\cite{salvalaglio2015molecular,zou2023driving} Indeed, crystalline nuclei may emerge from an intermediate dense liquid state, following the fluctuations in the local concentrations. It has been argued that the two-step process could be favorable in the nucleation of urea from aqueous solutions based on the analysis on the free energy profile.\cite{salvalaglio2015molecular} Here we utilize the AI-based analysis tool “State Predictive Information Bottleneck”\cite{wang2021state,mehdi2022accelerating} (SPIB) to investigate the nucleation mechanism through a reaction coordinate analysis. SPIB learns the reaction coordinate governing nucleation as a deep neural network, and has been previously employed for problems such as protein conformation change\cite{mehdi2022accelerating} as well crystal nucleation.\cite{zou2023driving} Our analysis suggests a prevailing kinetic preference of the system for the two-step mechanism, whose relation to the short and long-range interactions is also discussed.

\section*{Results}

\subsection*{Local molecular field theory} \label{LMF}

\begin{figure*}[h]
    \centering
    \begin{minipage}[b]{0.45\textwidth}
    \includegraphics[width=\textwidth]{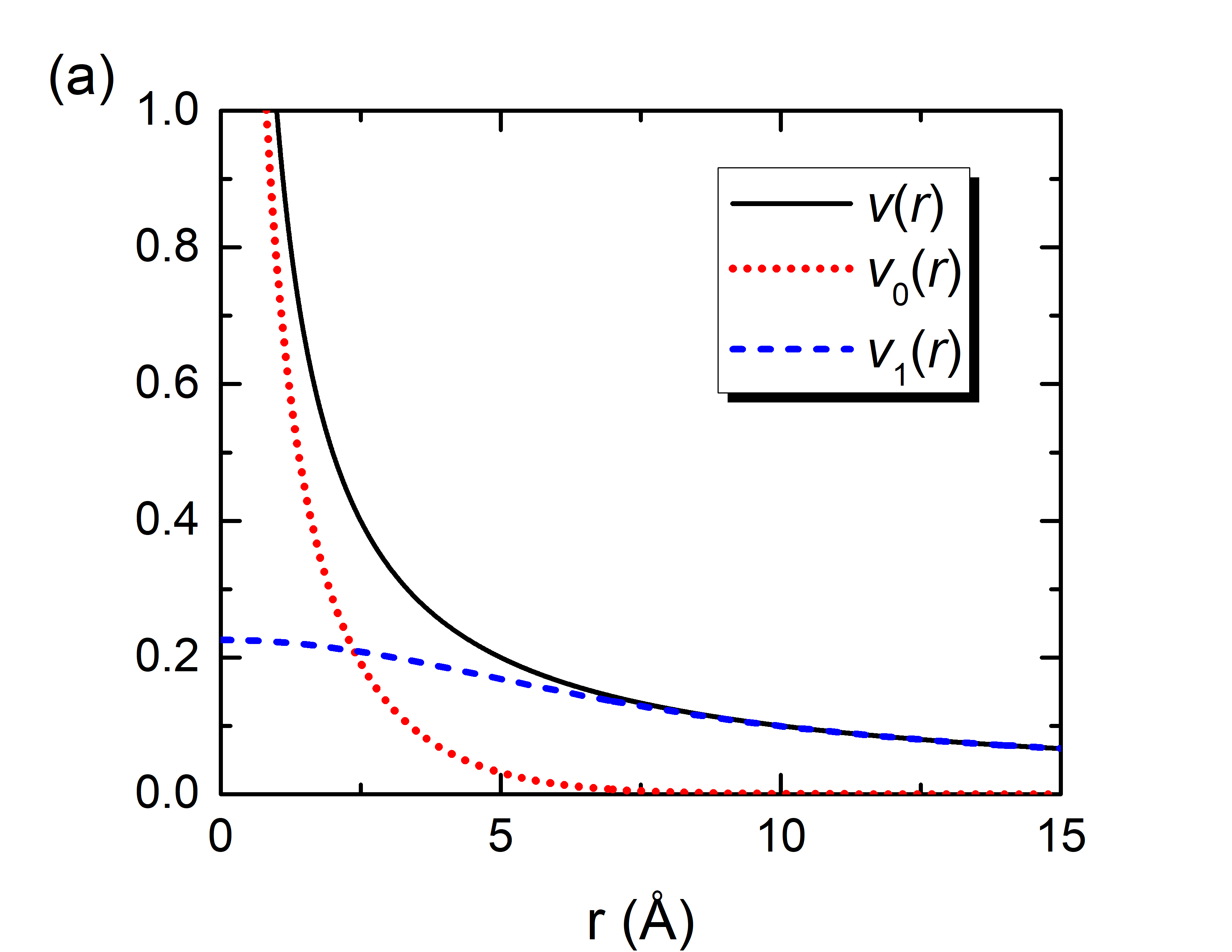}
    \end{minipage}
    \begin{minipage}[b]{0.45\textwidth}
    \includegraphics[width=\textwidth]{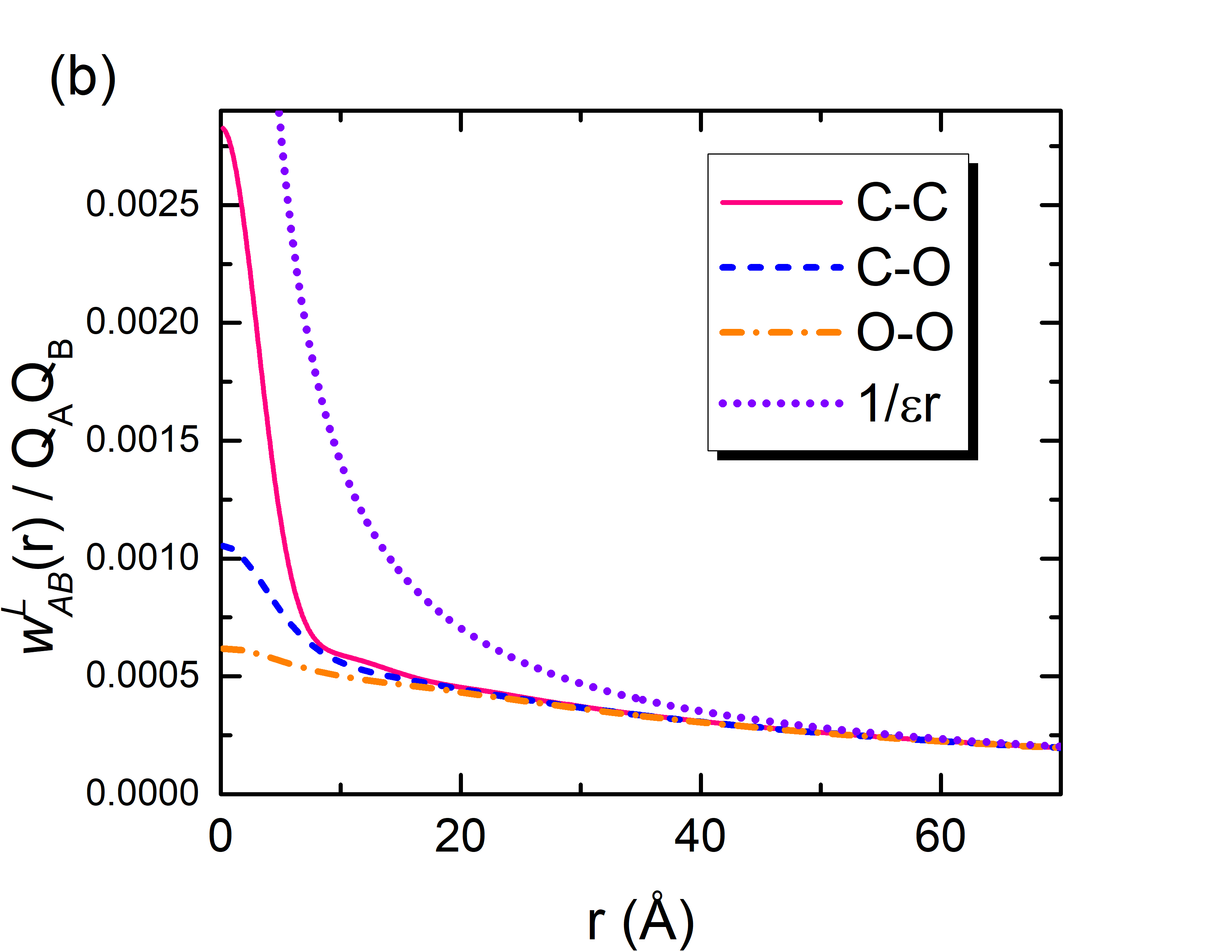}
    \end{minipage}
    \caption{ \textbf{The LMF descriptions for long-range interactions.} (a) Separation of the Coulomb potential $v(r)$ into the short-range component $v_0(r)$ and the long-range component $v_1(r)$, with the smoothing length $\sigma=5.0~\AA$. (b) Renormalized interactions $w_{AB}^{L}(r)$ among intermolecular carbon and oxygen sites of urea molecules solvated in SPC/E water. At large distances, they asymptotically approach the screened Coulomb potential (violet dotted line) predicted by the dielectric continuum theory.}
    \label{fig:LMF}
\end{figure*}

LMF theory separates the Coulomb interaction into short and long-range components, as shown in Fig. \ref{fig:LMF}(a),
\begin{equation}
v(r)\equiv\frac{1}{r}=\frac{\erfc{(r/\sigma)}}{r}+\frac{\erf{(r/\sigma)}}{r}\equiv v_0(r)+v_1(r),
\end{equation}
where erf and erfc are the error function and complementary error function. The long-range $v_1(r)$ arises from a unit Gaussian charge distribution of width $\sigma$, which contains only small wave vectors in reciprocal space and is therefore slowly varying in $r$-space. $v_1(r)$ approaches the Coulomb tail at distances greater than $\sigma$. The strong short-range $v_0(r)$ is the screened potential. It captures the forces from the full Coulomb potential at distances less than $\sigma$ while rapidly vanishes at distances greater than $\sigma$. The smoothing length $\sigma$ is chosen on the order of characteristic nearest-neighbor distances such that the short-range intermolecular correlations can be sufficiently captured by $v_0(r)$ and non-electrostatic interactions. In this article, we use $\sigma=5.0~\AA$ for model construction and MD simulations.

In earlier LMF work, they devised a Gaussian Truncated (GT) model\cite{remsing2011deconstructing,rodgers2009accurate} that replaces the point charge potential by the short-range $v_0(r)$. Since the hydrogen bonds in aqueous systems form as a result of the balance between the short-range electrostatic attraction between donor and acceptor charges and the repulsion of the overlapping Lennard-Jones cores, the GT model can be applied to the SPC/E water to reproduce the structural properties of bulk water to a high degree of accuracy.\cite{remsing2011deconstructing} The forces from the slowly varying long-range $v_1(r)$ cancel due to the uniformity of the bulk system.

Here, we take the GT model as the preliminary system in modelling the nucleation of urea in water. The urea molecules join the tetrahedral network of water by forming hydrogen bonds with water molecules,\cite{bandyopadhyay2014molecular} therefore the GT model is expected to largely retrieve the solvation structures in equilibrium. However, in the course of nucleation the system becomes highly nonuniform, encountering a situation where long-range interactions barely cancel, and a question arises whether the collective effects of long-range interactions can accumulate over the long timescale of nucleation.

To inquire into the above question, we utilize in our simulations the SS model, which takes the GT model as the reference system and renormalizes all the long-range interactions. In the SS model, we truncate all the Coulomb interactions between solvent-solvent sites and solute-solvent sites, retaining only the short-range electrostatic interactions $v_0(r)$. An effective solute-solute potential, as discussed in Ref. \onlinecite{gao2020short}, can be formulated based on LMF theory:
\begin{equation}
w_{\text{AB}}^{\text{SS}}(r)=w_{\text{ne}}(r)+Q_\text{A} Q_\text{B} v_0(r)+w_{\text{AB}}^{\text{L}}(r),
\label{eq:effective_potential}
\end{equation}
where $w_{\text{ne}}(r)$ is the nonelectrostatic interaction, $Q_\text{A}$ and $Q_\text{B}$ are the (partial) charges carried by solute atomic sites A and B, and $w_{\text{AB}}^{\text{SS}}(r)$ results from renormalizing all the long-range components of electrostatic interactions. $w_{\text{AB}}^{\text{SS}}(r)$ includes the averaged effects of the long-range electrostatics arising from the long-range solute-solvent and solvent-solvent interactions that are ignored in the reference GT system. The SS model takes advantage of the weak coupling between the short and long-range physics in charged and polar systems. It employs the short-range reference system (GT model) to capture the local molecular correlations, while simultaneously recovering the nonlocal effects through the renormalized interaction $w_{\text{AB}}^{\text{L}}(r)$ as a correction to the GT model. Fig. \ref{fig:LMF}(b) illustrates $w_{\text{AB}}^{\text{L}}(r)$ between the intermolecular carbon and oxygen sites of urea molecules solvated in SPC/E water. $w_{\text{AB}}^{\text{L}}(r)$ accounts for important solvent-mediated effects including particularly dielectric screening, as evidenced by its large distance asymptote $\frac{Q_\text{A} Q_\text{B}}{\epsilon r}$, which is the macroscopic limit of dielectric screening.

In previous work, the effectiveness of the SS model was demonstrated in modelling the pairing of monovalent ions, multivalent ions, and nucleobases.\cite{gao2020short,gao2023local} Our motivation is that it should also be able to capture the correlations between urea molecules in the process of nucleation from aqueous solutions. By then comparing the simulation results from the SS model to that from the GT model, we should be able to disentangle the roles of the short and long-range interactions in such complex dynamical processes.

The full model treats the long-range interactions of all charged sites using Ewald sums and related methods,\cite{allen2017computer,belhadj1991molecular,essmann1995smooth} but the indirect “black-box" nature of the lattice sum algorithms hampers the development of simple physical pictures directly incorporating dielectric screening between solutes. The SS model optimizes the model complexity on different length scales by encoding the essential long-range effects in the renormalized solute-solute interaction, while retaining the local correlations in the short-range reference system. 

In the following context we combine the SS model with WTmetaD, an advanced sampling method proven effective in the computational studies of nucleation.\cite{giberti2015metadynamics}

\subsection*{Nucleation of urea in water}

To investigate phase transitions occurring in aqueous solutions of urea at ambient conditions, the well-tempered metadynamics\cite{barducci2008well} approach is utilized. This methodology incorporates a time-dependent potential through the deposition of Gaussian kernels as a function of the biased collective variable, thereby depressing frequently visited configurations and enhancing exploration of the phase space. Following the previous work by Piaggi et al. \cite{ piaggi2018predicting}, we consider the approximate pair orientational entropy as the collective variable being biased. It is defined as
\begin{multline}
S(r,\theta)=-\pi\rho k_\text{B}\int_{0}^{\infty}\int_{0}^{\pi}[g(r,\theta)\ln{g(r,\theta)}-g(r,\theta)+1]\\
\times r^2\sin{\theta}drd\theta,
\end{multline}
where $\rho$ is the number density of solute molecules, $k_\text{B}$ is Boltzmann constant, $r$ is the intermolecular distance, $\theta$ is the intermolecular angle between characteristic vectors of solute molecules, and $g(r, \theta)$ is the correlation function.\cite{prestipino2004entropy} The pair orientational entropy comes from the leading term in the expansion of the excess entropy in the liquid state theory.\cite{prestipino2004entropy} It proved useful in searching for polymorphs in simulations of urea and naphthalene in vacuum.\cite{ piaggi2018predicting} In the case of aqueous solutions, it is representative of the contribution of solute-solute correlations to the system entropy. The characteristic vectors for urea are chosen to be along the direction of the carbonyl group (which is also the direction of the dipole moment) and the direction connecting the two nitrogen atoms, and the resultant intermolecular angles are denoted as $\theta_1$ and $\theta_2$, respectively, as illustrated in Fig. 2(b). In our simulations, biasing $S(r,\theta_1)$ and $S(r,\theta_2)$ enables the enhancement of fluctuations in both spatial and orientational correlations without requiring empirical knowledge of crystal structures. To facilitate the identification of different phases of urea, we use the averaged intermolecular angles, $\bar{\theta}_1$ and $\bar{\theta}_2$, the detailed expression of which are provided in Supplementary Note 1. In the SPIB analysis discussed in the next section, we also include the second moments $\mu^2_{\theta_1}$ and $\mu^2_{\theta_2}$.

We conducted three independent 300 ns runs for each model in the WTmetaD simulations performed with the GT model, the SS model, and the full model. Fig. 3 (a)-(c) illustrate the time series of $\bar{\theta}_1$ and $\bar{\theta}_2$ for representative trajectories for the three models. The system stays in the liquid-like state when $\bar{\theta}_1$ and $\bar{\theta}_2$ both fluctuate around $1.0$. These two variables dropping below certain thresholds indicates the emergence of crystal nuclei. As shown in Fig. 3 (a), $\bar{\theta}_1$ being smaller than $\bar{\theta}_2$ indicates that the system is in form I, the most stable crystal structure of urea at ambient conditions.\cite{giberti2015insight, swaminathan1984crystal} The crystal state with $\bar{\theta}_2$ larger than $\bar{\theta}_1$ corresponds to form B.\cite{piaggi2018predicting} In Fig. 3 (d)-(f), we computed the equilibrium free energy surfaces (FESs) in the $(\bar{\theta}_1,\bar{\theta}_2)$ space by reweighting the metadynamics simulations.\cite{tiwary2015time} Three basins are found corresponding to the liquid-like state (cyan star), form I (blue circle), and form B (pink triangle). It is shown that, in comparison to those of form B, form I possesses a remarkably lower free energy barrier when transitioning from/to the liquid-like state, as well as a lower free energy minimum. It is currently unknown whether form B arises from artifacts due to the use of the generalized Amber force field\cite{case2005amber} or if it has yet to be experimentally observed, but based on the information from our FESs, it is unlikely to exist at ambient conditions.  Urea is reported to have rich polymorphism including many high-pressure and high-temperature products.\cite{polymorphism1,lamelas2005raman,dziubek2017high,roszak2017giant} Here we focus on the most probable transitions at ambient conditions and do not attempt to improve the efficiency of polymorph search. It should also be noted that the free energy of the crystal states is influenced by the finite size effect associated with the chemical potential, which can be corrected analytically using CNT\cite{salvalaglio2015molecular} or avoided by the constant chemical potential MD simulation method,\cite{perego2015molecular,karmakar2019molecular} but this is beyond the scope of the current work.

As shown in Fig. 3, the GT model, the SS model, and the full model lead to qualitatively similar FESs. From such a qualitative analysis, one could be tempted to support a physical picture that the nucleation processes are primarily driven by short-range interactions, a characteristic commonly captured by the three models under consideration. However, the situation is more subtle when we further quantify the possible effects of long-range interactions. To do so, we estimate for all 3 models the free energy difference between the liquid-like state and experimentally stable form I polymorph, defined as
\begin{equation}
\Delta G=G_\text{I}-G_\text{L}=\frac{1}{\beta}\log\frac{P_\text{L}}{P_\text{I}},
\label{eq:FED}
\end{equation}
where $\beta$ is the inverse temperature, and $P_\text{L}$ and $P_\text{I}$ are the probabilities for the system to appear in the vicinity of free energy minima of the liquid-like basin and the form I basin (with the landmarks labeled in  Fig. 3 (d)-(f)). For the biased WTmetaD simulations, $P_\text{L}$ and $P_\text{I}$ are computed with a reweighting protocol.\cite{tiwary2015time} Fig. 4 presents the free energy differences computed for the three models. For each model the result is averaged from three independent 300 ns trajectories, and the error bar is computed accordingly. The comparison of free energy differences reveals that the error bars of the full model and the SS model overlap, whereas the GT model exhibits a free energy difference higher than those of the other two models by $\sim 100~\text{kJ}~\text{mol}^{-1}$. This indicates that long-range interactions play a role in further stabilizing the main crystal state in addition to the effects attributed to short-range physics. Moreover, the inclusion of renormalized solute-solute interactions in the SS model effectively captures the impact of these long-range interactions. A purely short ranged model is thus incapable of describing the thermodynamics of urea nucleation in water.

\subsection*{SPIB analysis for reaction coordinate of nucleation}

Previous work has suggested that urea nucleation in aqueous solutions follows a two-step nucleation process \cite{salvalaglio2015molecular} which may take place on different time scales, highlighting the need for the analysis on metastability. Thus, in addition to describing the free energy landscape, the extent to which the LMF-based models can recover the metastability of nucleation is also of interest. Here we use State Predictive Information Bottleneck\cite{wang2021state,mehdi2022accelerating,ribeiro2018reweighted} to conduct such an analysis by constructing the reaction coordinate characterizing nucleation. SPIB seeks to learn RCs from high-dimensional trajectory data generated by MD simulations. The learned RC is constructed as a past-future information bottleneck to accurately predict the future state of the system with minimal information from the past. SPIB assumes that the system's dynamics is comprised of motions between metastable states, with the state-to-state transition time being slower than the intra-state relaxation time. Through a self-consistent procedure, SPIB automatically partitions the high-dimensional configuration space into such metastable states. The level of coarse-graining is determined by a hyper-parameter called the time delay $\Delta t$. Predictions are made for the system's future state after time $\Delta t$, excluding fluctuations within each metastable state that occur faster than transitions between states. For a 2-state system, this has been shown to be equivalent to the committor.\cite{wang2021state} Learning the SPIB allows us to separate timescales and analyze distinct steps in nucleation processes.

The three independent 300 ns trajectories simulated with the SS model are subjected to SPIB analysis. To characterize the input information of the high-dimensional trajectory data, a library of collective variables is employed including $S(r,\theta_1)$, $S(r,\theta_2)$, $\bar{\theta}_1$, $\bar{\theta}_2$, $\mu^2_{\theta_1}$, $\mu^2_{\theta_2}$, and a set of variables derived from the coordination number of urea, namely $N_{8+}$, $N_{11+}$, which are the populations of molecules with coordination numbers greater than 8 and 11, and the second moment of coordination numbers $\mu^2_c$. The analytical function for computing the coordination number is described in Supplementary Note 2. The coordination number-based variables are indicative of the size of clusters and the fluctuations in local concentrations. 

Fig. 5(a) shows the one-dimensional free energy along $\bar{\theta}_1$. Since we focus on the dominant transitions between the liquid-like state and the primary crystal state (form I) occurring at ambient conditions, the limited number of transitions from and to form B are excluded from our analysis. Thus Fig. 5(a) shows only the liquid-like state and the primary crystal state. We emphasize here that $\bar{\theta}_1$ is sufficient to distinguish between the two states with a clear barrier located at $\bar{\theta}_1\sim0.95$. Given that the dipole moments along the C-O axis of neighboring urea molecules are either parallel or antiparallel in form I, $\bar{\theta}_1$ naturally characterizes the degree of crystallinity.  Fig. 5(b) plots the reweighted FES in the $(\bar{\theta}_1,N_{8+})$ space, in which it is explicitly shown that, as $ N_{8+}$ can significantly fluctuate in it, the liquid-like basin contains configurations of dense liquid clusters.

The state labels learnt by SPIB are projected into the $(\bar{\theta}_1,N_{8+})$ space in Fig. 5(c). Three states are identified by SPIB using $\Delta t=300$ ps, which is chosen on the scale of the minimal time resolution for distinguishing the metastable states under consideration. State $\alpha$ marked in Fig. 5(c) and colored mustard exactly occupies the liquid-like basin and is therefore recognized as the liquid-like state. A significant portion of the crystal basin is occupied by State $\beta$ colored fuchsia. This state includes the configurations with small to moderate-sized crystal nuclei in the experimental stable form I and coincides with the most stable regions in the crystal basin. State $\gamma$ colored orange corresponds to the configurations with crystal nuclei of large sizes, the accessibility to which is limited by the finite size effects associated with the chemical potential,\cite{karmakar2019molecular,salvalaglio2015molecular} therefore they are less frequently visited by the system and labeled differently from State $\beta$. In Fig. 6, with SPIB one can differentiate between State $\alpha$ and State $\beta$ with the free energy barrier for all three models. However, in contrast to the other two models, State $\gamma$ for the GT model is barely detected. This can be attributed to the GT model ignoring the long-range interactions, which play a more significant role when the system is highly nonuniform at large $\bar{\theta}_1$ and $N_{8+}$. Thus, once again, similar to the conclusion drawn in Fig. \ref{fig:FED} where SS model approximated full model far better than the GT model for getting free energy differences, we find that careful treatment of long-range interactions as done in the SS model is needed to account for metastability during urea nucleation.

We now examine the transition between these three metastable states, ignoring those within as irrelevant for the particular choice of time-delay. Spanning over a broad range of $N_{8+}$, state $\alpha$ (the liquid-like state) exhibits large fluctuations in the local concentrations or the size of dense clusters, while the onset of crystal formation occurs with the transition from State $\alpha$ to State $\beta$. Since intra-state fluctuations in SPIB are per construction faster than inter-state, this shows that, starting from the liquid state, the fluctuations in the local concentrations take place on a much shorter timescale than the fluctuations in the structural order. This is to be contrasted with CNT, which requires the evolution of crystallinity being simultaneous with the changes in the cluster size. Thus the nucleation processes we observe for urea in aqueous solutions cannot be adequately accommodated within this framework. Indeed, the state labels plotted in the SPIB-learnt RC space (shown in Fig. 5(d)) are topologically very similar to those in the $(\bar{\theta}_1,N_{8+})$ space, implying that the system evolves along two RCs relatively independent of one another, with one mainly related to the cluster size or local concentrations and the other mainly related to the structural orders.

To further enquire whether the fluctuations in local concentrations is a prerequisite for the nucleation of crystals, we consider the distribution of $N_{8+}$ when the transitions between State $\alpha$ and State $\beta$ occur, which is detailed in Supplementary Note 3. The interconversion between the two states can take place over a wide range of $N_{8+}$, but the system exhibits a tendency to have the dense clusters grown to certain sizes before transitioning to the crystal state. A 50 ns unbiased simulation starting from the liquid state yields an average $N_{8+}$ value of 4.3 with a standard deviation of 1.77. By comparison, biased simulations show that $N_{8+}$ values within the three-sigma limits of the above liquid state $N_{8+}$ distribution occur in only $8.7\%$ of the total 46 transitions, namely $8.7\%$ of the transitions between the liquid-like basin and the primary crystal basin take place directly from or to the liquid state, and the other transitions are mediated by the dense liquid state. This suggests that the nucleation events in our simulations predominantly proceed through a two-step process. Since similar results of $N_{8+}$ are obtained for the GT model and the full model (shown in Supplementary Figure 1), the intricate nucleation mechanism leading to the two-step process is largely captured by the short-range interactions. However, to fully recover the detailed dynamics when the system is highly nonuniform, one needs to take into account the effects of long-range interactions, which can be done using the SS model as we have illustrated by Fig. 5(c) and 6.

\section*{Discussion}
The presence of solvents in aqueous solutions introduces significant challenges in efficiently sampling the nucleation events, assessing the solvent-mediated effects due to short and long-range interactions, and comprehending the underlying nucleation mechanisms. The recent advancements in machine learning have significantly contributed to the development of molecular models for aqueous systems. However, conventional machine learning models are limited to short-range interactions, and an effective handling of long-range interactions is in demand for scenarios where such interactions play important roles.\cite{lambros2021general,yue2021short,gao2022self} In this work, we have investigated the nucleation processes of urea in water at ambient conditions, by using the LMF-based models in WTmetaD simulations. The SS model was introduced by LMF theory to construct effective solute-solute potentials by renormalizing long-range components of electrostatic interactions. It is used as a correction to the reference GT model, which considers only short-range interactions. The comparison between the results drawn from the two models provides us with insights into the different roles of short and long-range interactions in the processes of urea nucleation in water. In WTmetaD simulations, systems simulated with the GT model, the SS model and the full model can explore the same regions of the state space, sampling the liquid phase, the dense liquid phase, and the primary crystal phase at ambient conditions. Since the short-range interactions are common to the three models, it is inferred that the nucleation processes are primarily driven by the short-range physics, which encompasses a combination of hydrogen bonding, molecular packing, and van der Waals attractions. The effects of long-range interactions can be manifested in various subtle aspects such as the association of the solute molecules, the attachment of mobile solutes to crystal nuclei, and the chemical potential that involves the correlations between distant solute molecules, etc., which are hard to quantify separately in the long duration of nucleation. The applications of the SS model and WTmetaD facilitate the determination of the collective response of the system to these effects. An in-depth analysis of the free energy difference reweighted from WTmetaD highlights the significant contribution of long-range interactions towards stabilizing the main crystal state, which can be well captured by the SS model in a simple and physically suggestive framework.

Our simulation results are further analyzed by the AI-based approach SPIB, which allows us to approximate the RCs describing the relevant slow dynamics of state-to-state transitions and to make time scale decompositions for the nucleation processes. It is shown that, starting from the liquid state, the system can evolve along the 2 dimensions of the SPIB-learnt RCs, one induced by the fluctuations in the local concentrations and the other induced by the fluctuations in the structure orders, and these two types of fluctuations occur on distinctly different timescales. We have demonstrated that the slow progression of crystal formation is predominantly preceded by the fast fluctuations in the local concentrations, which is indicative of a two-step nucleation process. Through the comparison among the metastable states detected by SPIB for the GT model, the SS model, and the full model, we have confirmed that the two-step mechanism is largely captured by the short-range interactions common to the three models, while the inclusion of long-range interactions is indispensable for describing the detailed metastability during urea nucleation in water.

\section*{Methods}

\subsection*{Short solvent model}

The renormalized solute-solute long-range interaction $w_{\text{AB}}^{\text{L}}(r)$ of the SS model in Eq. \ref{eq:effective_potential} is calculated with the following expression,
\begin{equation}
\begin{split}
w_{\text{AB}}^{\text{L}}(r) = & \quad Q_\text{A} Q_\text{B} v_1(r) \\
 & +\frac{1}{2} \int d\bm{\mathrm{r}^\prime} (\rho_\text{A}^q(r^\prime)+\rho_{0,\text{A}}^q(r^\prime)) Q_\text{B} v_1(|\bm{\mathrm{r}^\prime} - \bm{\mathrm{r}}|) \\
 & +\frac{1}{2} \int d\bm{\mathrm{r}^\prime} (\rho_\text{B}^q(r^\prime)+\rho_{0,\text{B}}^q(r^\prime)) Q_\text{A} v_1(|\bm{\mathrm{r}^\prime} - \bm{\mathrm{r}}|) \\
 & +\frac{1}{2} \iint d\bm{\mathrm{r}^\prime} d\bm{\mathrm{r}^{\prime\prime}} (\rho_\text{A}^q(r^\prime)\rho_{0,\text{B}}^q(r^{\prime\prime}) \\
 & \qquad +\rho_\text{B}^q(r^\prime)\rho_{0,\text{A}}^q(r^{\prime\prime})) v_1(|\bm{\mathrm{r}^{\prime\prime}} - \bm{\mathrm{r}^\prime}|)
\end{split}
\end{equation}
where $\rho_\text{A}^q(r)$ and $\rho_{0,\text{A}}^q(r)$ are the singlet solvent charge density induced by the hydration of interaction site A in the full and the GT systems (denoted with subscript 0), respectively.

\subsection*{State Predictive Information Bottleneck}
Belonging to the RAVE\cite{ribeiro2018reweighted,wang2019past,wang2020machine} family of methods, SPIB focuses on the relevant slow dynamics of the system and constructs RCs to predict transitions between metastable states. The extent of coarse-graining in time is controlled by a tunable hyper-parameter, the time delay $\Delta t$, and all the fast modes are filtered out accordingly. In this manner, we decompose the timescales of the possibly different stages in the transitions from the liquid phase to the main crystal phase.

\subsection*{Simulation setup}

We considered a system composed of 300 urea molecules described by the generalized Amber force field\cite{case2005amber} and 3057 SPC/E\cite{berendsen1987missing} water molecules. Lorentz-Berthelot mixing rules\cite{lorentz1881ueber} were used to determine the cross-interaction parameters. All the simulations were performed with a time step of 2.0 fs in the isothermal isobaric (NPT) ensemble at $T$ = 300 K and $P$ = 1 atm, maintained by a Nos\'{e}-Hoover thermostat\cite{nose1984unified} and barostat,\cite{nose1983constant} with damping constants of 0.1 and 1.0 ps, respectively. The pressure correction due to the truncated Coulomb
interactions of the solvents was applied to the GT model and the SS model.\cite{rodgers2009accurate} Lennard-Jones interactions were computed up to the cutoff of 10~\AA. The short-range GT term $v_0$ was realized by tabulating the potential up to 10~\AA. For the full model, long-range Coulomb interactions were evaluated using particle-particle particle-mesh method with a real space cutoff of 10~\AA. For the SS model, only long-range effective solute-solute interactions $w_{\text{AB}}^{\text{L}}(r)$ were evaluated using particle-particle particle-mesh method. The systems were equilibrated for 1 ns before metadynamics simulations. For each model, three independent 300 ns WTmetaD production runs were generated, with the height of the Gaussian deposition set to 2 $k_\text{B} T$, the width of the Gaussian set to 0.2, which approximates the typical thermal fluctuation of the collective variable being biased, and the bias factor set to 100. We utilized the LAMMPS package\cite{plimpton1995fast} patched with PLUMED 2.7\cite{tribello2014plumed} in carrying out our simulations.

\section*{Data availability}
The input files necessary to reproduce the simulations done in this work are available on PLUMED-NEST at https://www.plumed-nest.org/eggs/23/034/.

\section*{Acknowledgement}
This work is entirely funded by the US Department of Energy, Office of Science, Basic Energy Sciences, CPIMS Program, under Award DE-SC0021009. We are grateful to NSF ACCESS Bridges2 (project CHE180053) and University of Maryland Zaratan High-Performance Computing cluster for enabling the work performed here. We also thank Dedi Wang for fruitful discussions.

\section*{Author contributions}
R.Z., Z.Z., J.D.W., and P.T. designed research; R.Z. performed research and analyzed data; and R.Z., Z.Z., J.D.W., and P.T. wrote the paper.

\section*{Competing interests}
The authors declare no competing interests.

\hfill \break
\normalem
\textbf{References}
\bibliographystyle{naturemag}
\bibliography{references}

\section*{Figures}

\begin{figure*}[th]
    \centering
    \includegraphics[width=0.7\textwidth]{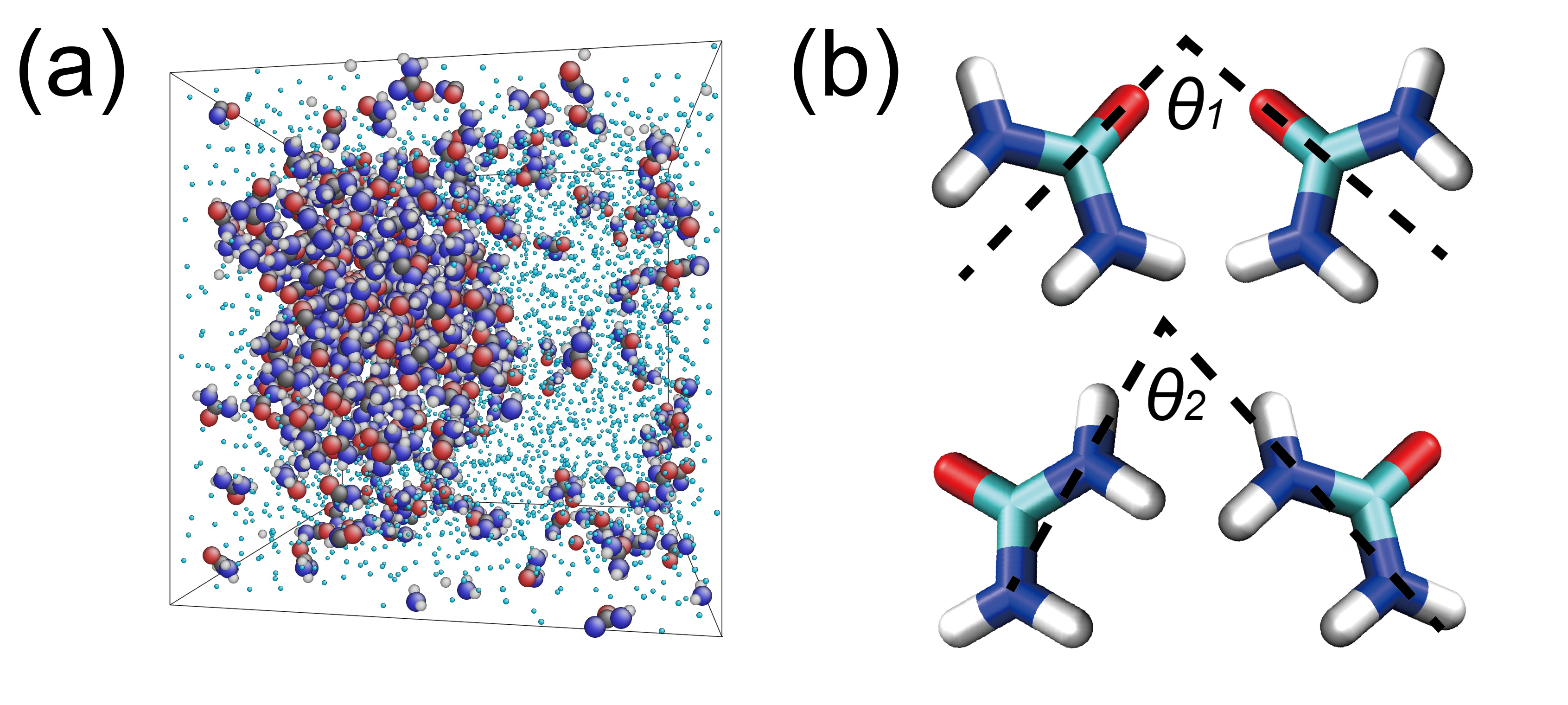}
    \caption{ \textbf{Illustration of the simulation system.} (a) Snapshot of a urea cluster in aqueous solution. For water molecules only oxygen sites are displayed. (b) The characteristic vectors and the corresponding intermolecular angles for two neighboring urea molecules.}
    \label{fig:urea}
\end{figure*}

\begin{figure*}[t]
    \centering
    \begin{minipage}[b]{0.3\textwidth}
    \includegraphics[width=\textwidth]{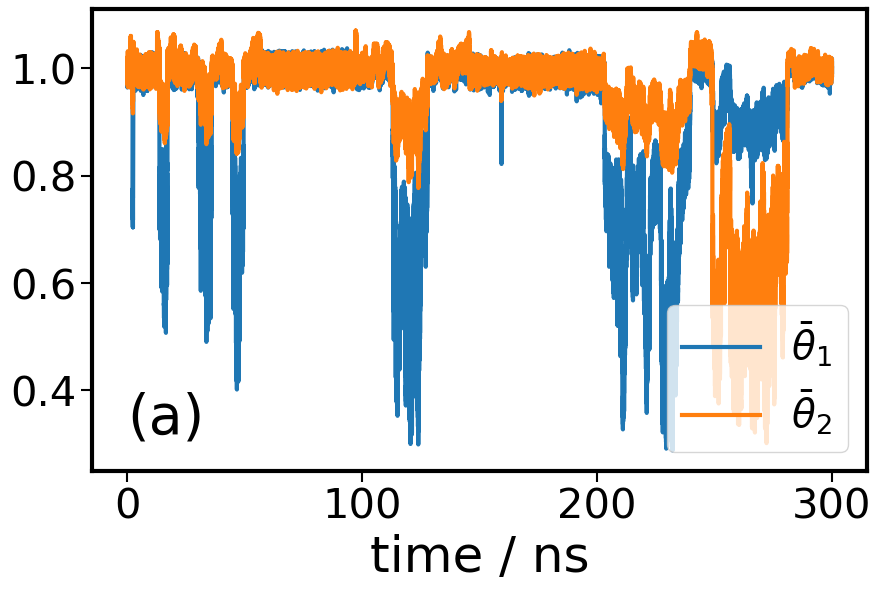}
    \end{minipage}
    \begin{minipage}[b]{0.3\textwidth}
    \includegraphics[width=\textwidth]{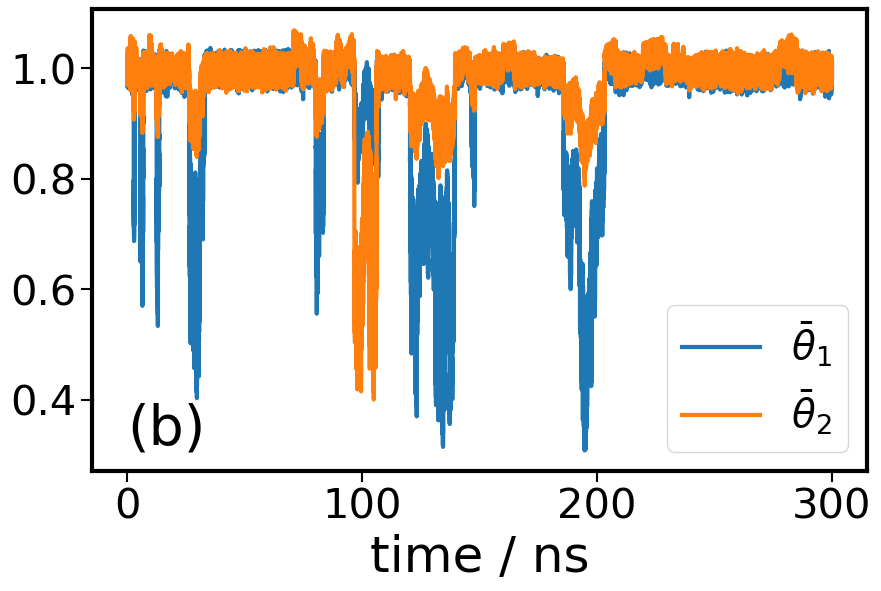}
    \end{minipage}
    \begin{minipage}[b]{0.3\textwidth}
    \includegraphics[width=\textwidth]{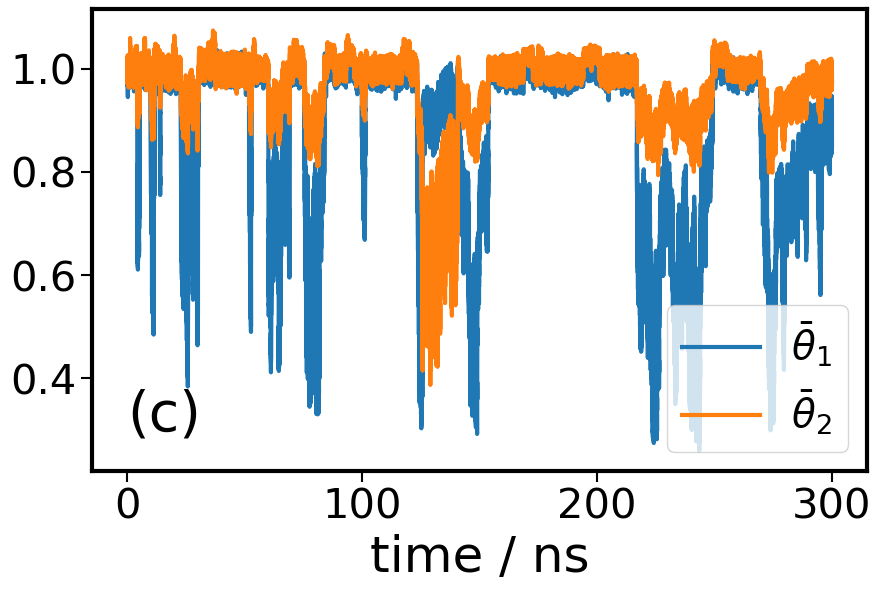}
    \end{minipage}
    \begin{minipage}[b]{0.3\textwidth}
    \includegraphics[width=\textwidth]{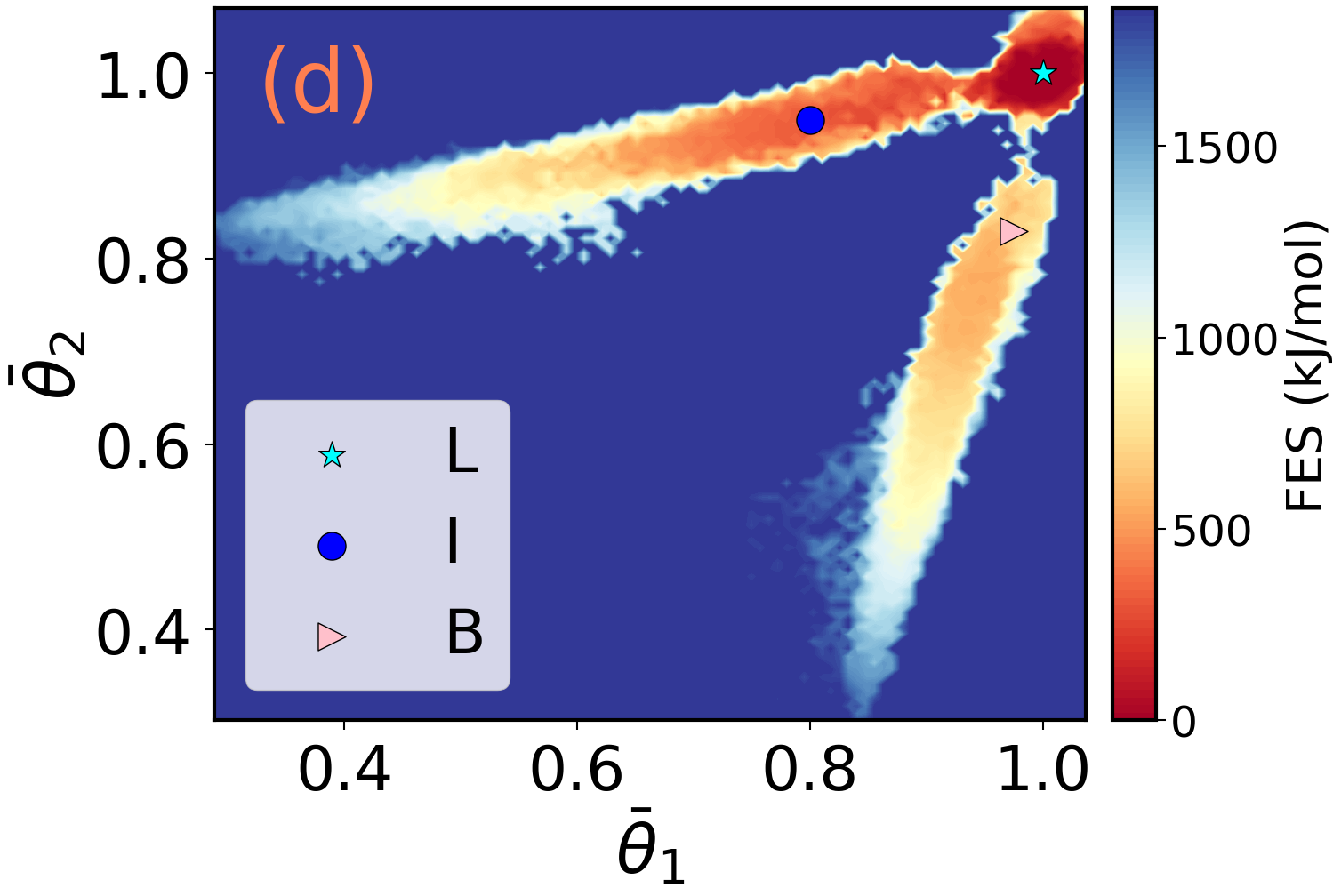}
    \end{minipage}
    \begin{minipage}[b]{0.3\textwidth}
    \includegraphics[width=\textwidth]{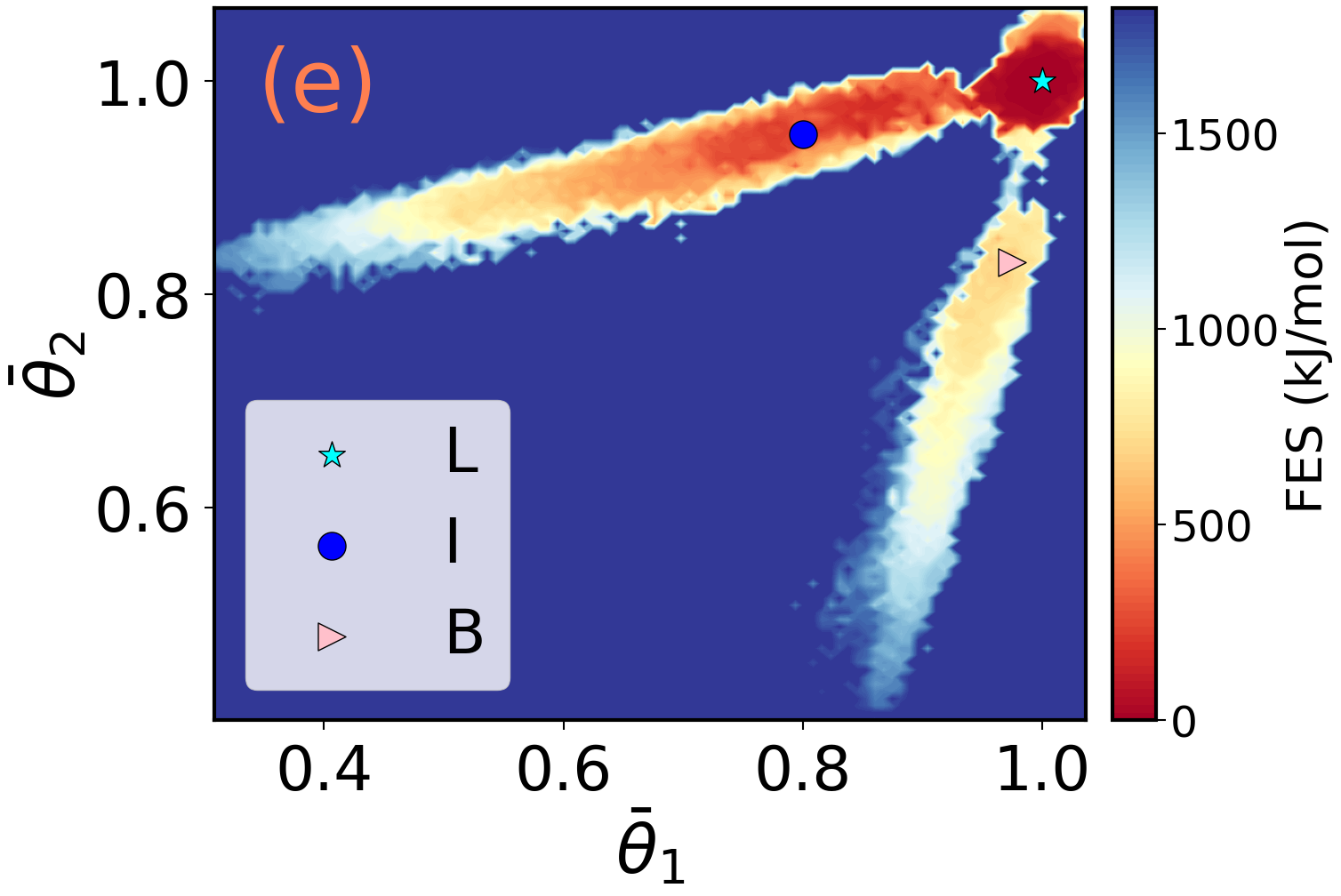}
    \end{minipage}
    \begin{minipage}[b]{0.3\textwidth}
    \includegraphics[width=\textwidth]{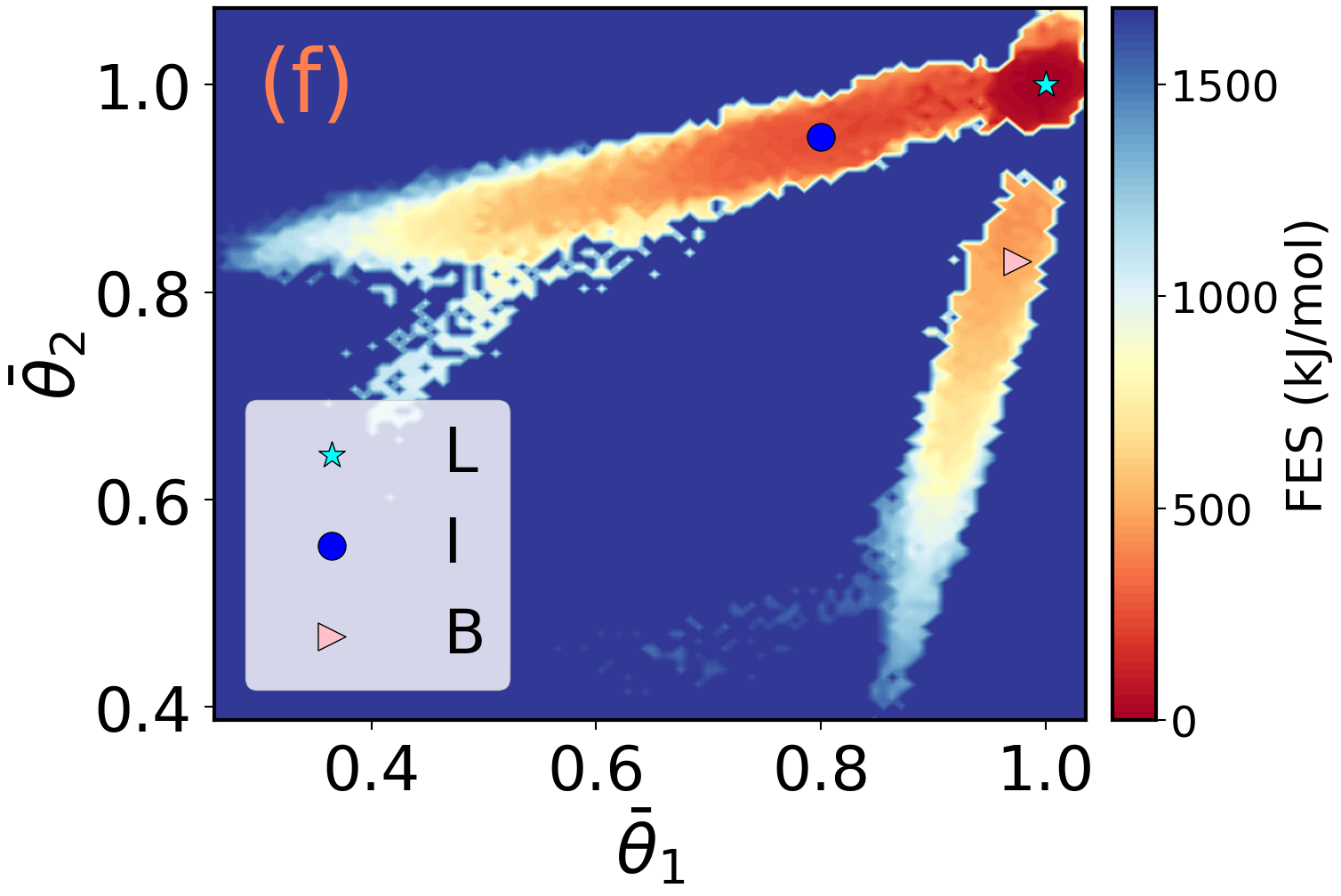}
    \end{minipage}
    \caption{\textbf{Time series and free energy surfaces.} In the top row are the time series of averaged intermolecular angles $\bar{\theta}_1$ (orange) and $\bar{\theta}_2$ (blue) extracted from the representative trajectories of WTmetaD simulations (biasing the pair orientational entropies $S(r,\theta_1)$ and $S(r,\theta_2)$) for (a) the GT model, (b) the SS model, and (c) the full model. In the bottom row are the free energy surfaces reweighted from the corresponding WTmetaD simulations performed with (d) the GT model, (e) the SS model, and (f) the full model.}
    \label{fig:compare}
\end{figure*}

\begin{figure*}[th]
    \centering
    \begin{minipage}[b]{0.4\textwidth}
    \includegraphics[width=\textwidth]{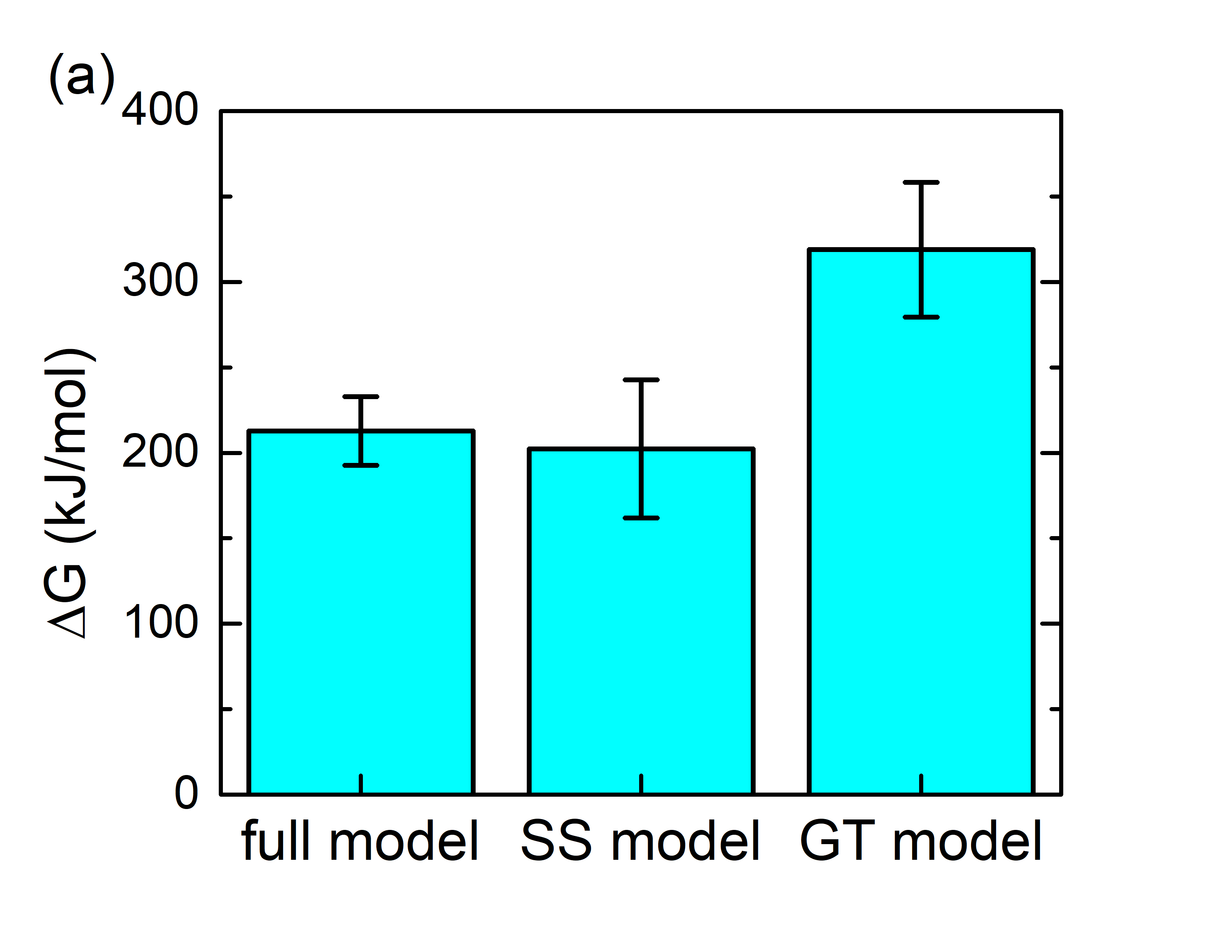}
    \end{minipage}
    \begin{minipage}[b]{0.4\textwidth}
    \includegraphics[width=\textwidth]{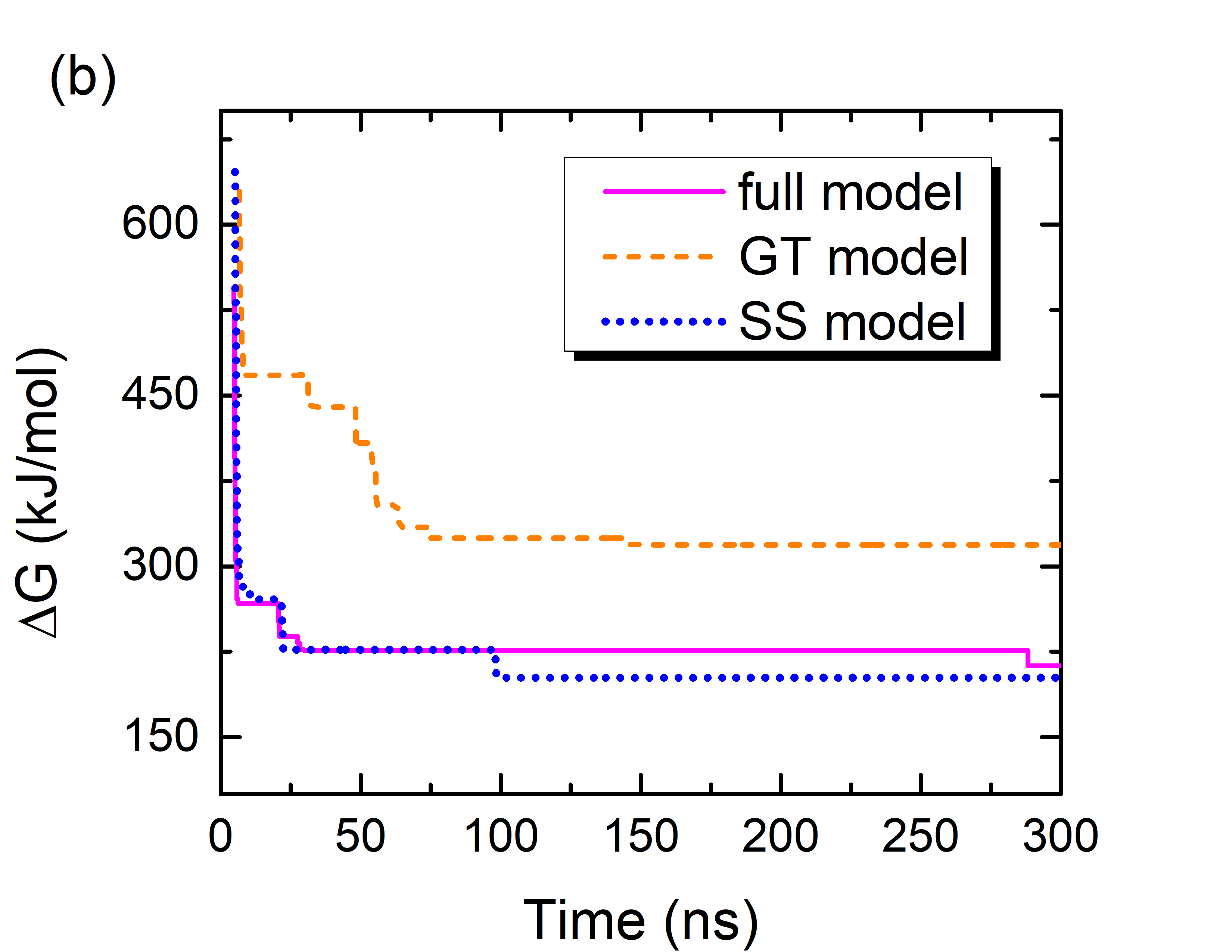}
    \end{minipage}
    \caption{\textbf{Free energy differences comparison.} Free energy differences between the liquid-like state and form I computed with the GT model, the SS model, and the full model. The error bar (standard deviation) for each model in (a) is computed over three independent production runs. (b) shows the convergence of averaged free energy differences with simulation time.}
    \label{fig:FED}
\end{figure*}

\begin{figure*}[hb]
    \centering
    \begin{minipage}[b]{0.38\textwidth}
    \includegraphics[width=\textwidth]{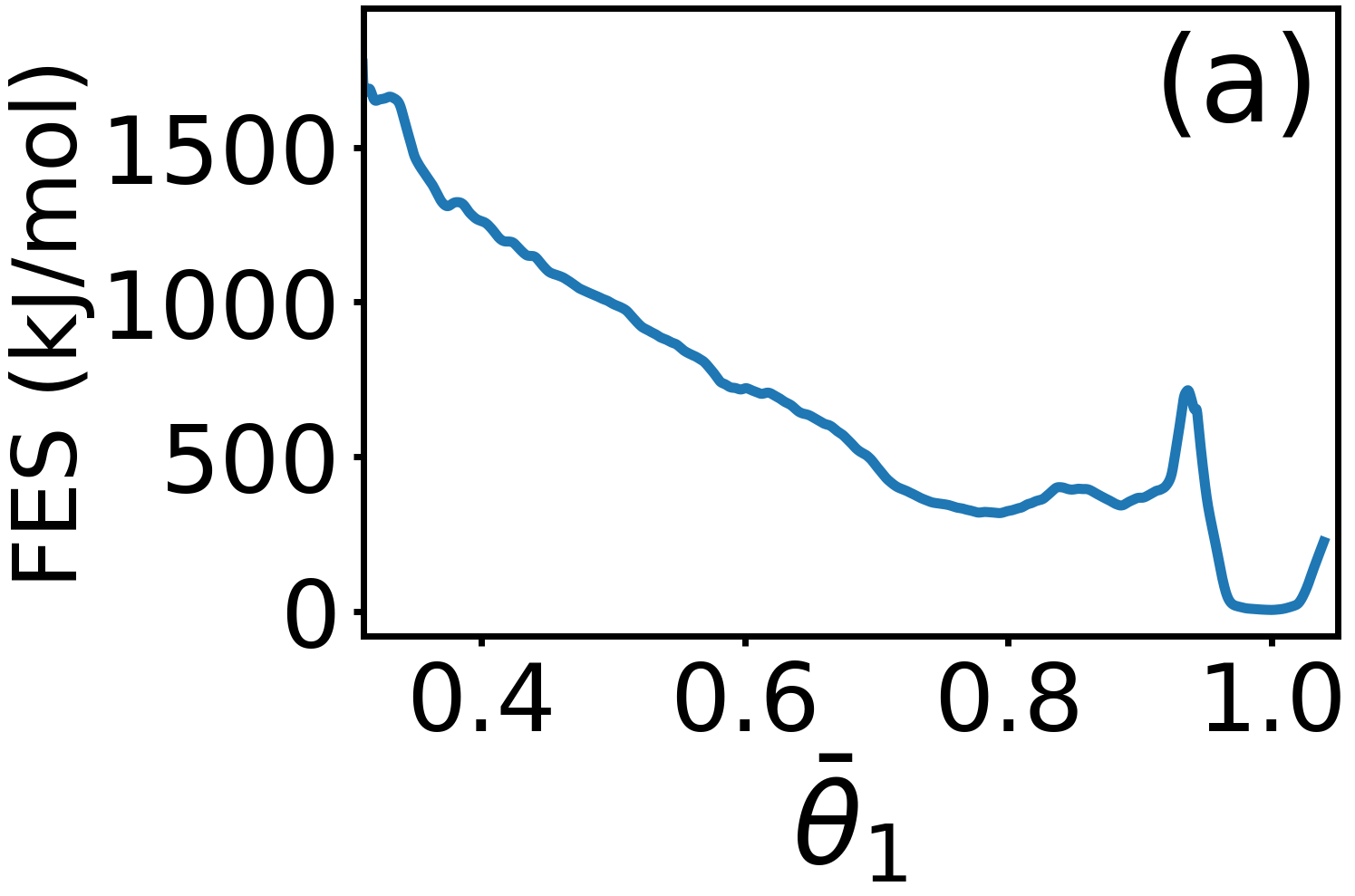}
    \end{minipage}
    \begin{minipage}[b]{0.38\textwidth}
    \includegraphics[width=\textwidth]{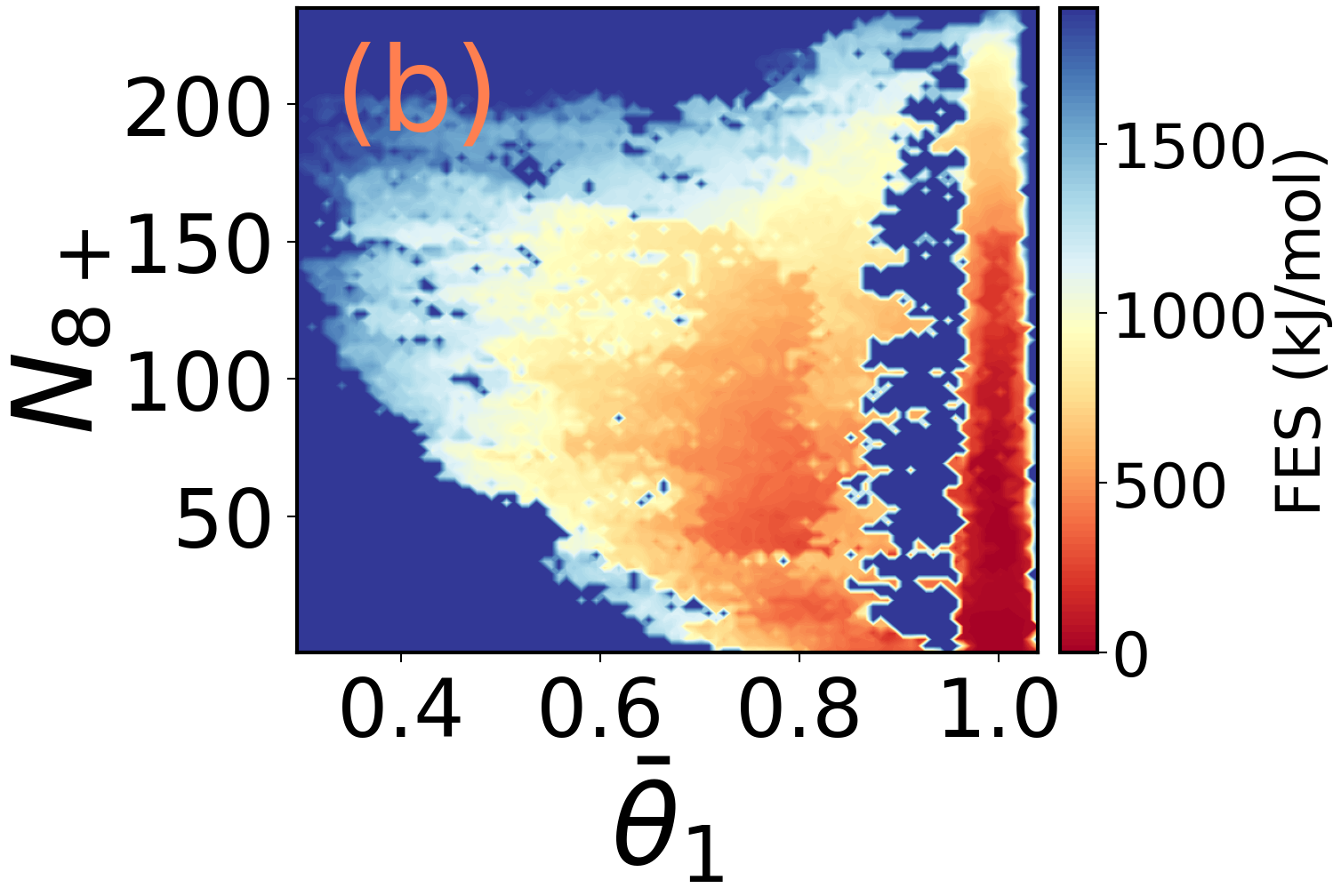}
    \end{minipage}
    \begin{minipage}[b]{0.38\textwidth}
    \includegraphics[width=\textwidth]{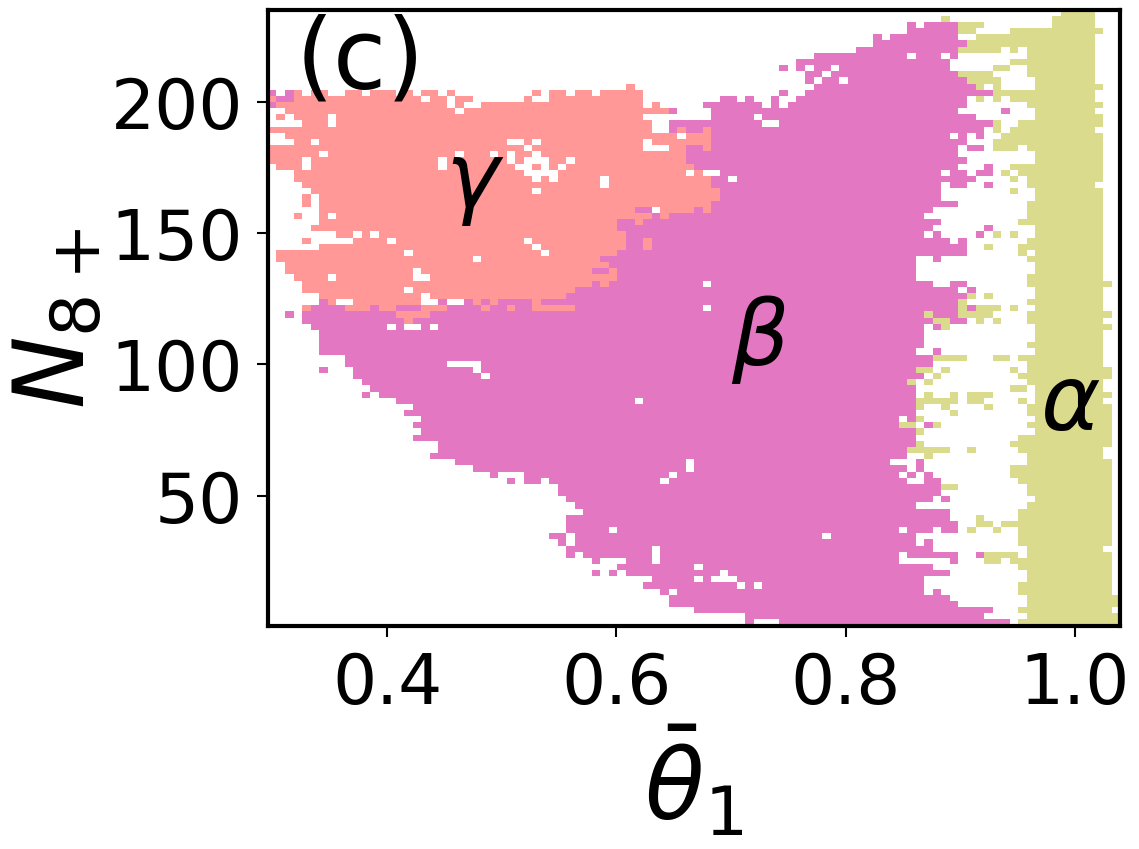}
    \end{minipage}
    \begin{minipage}[b]{0.38\textwidth}
    \includegraphics[width=\textwidth]{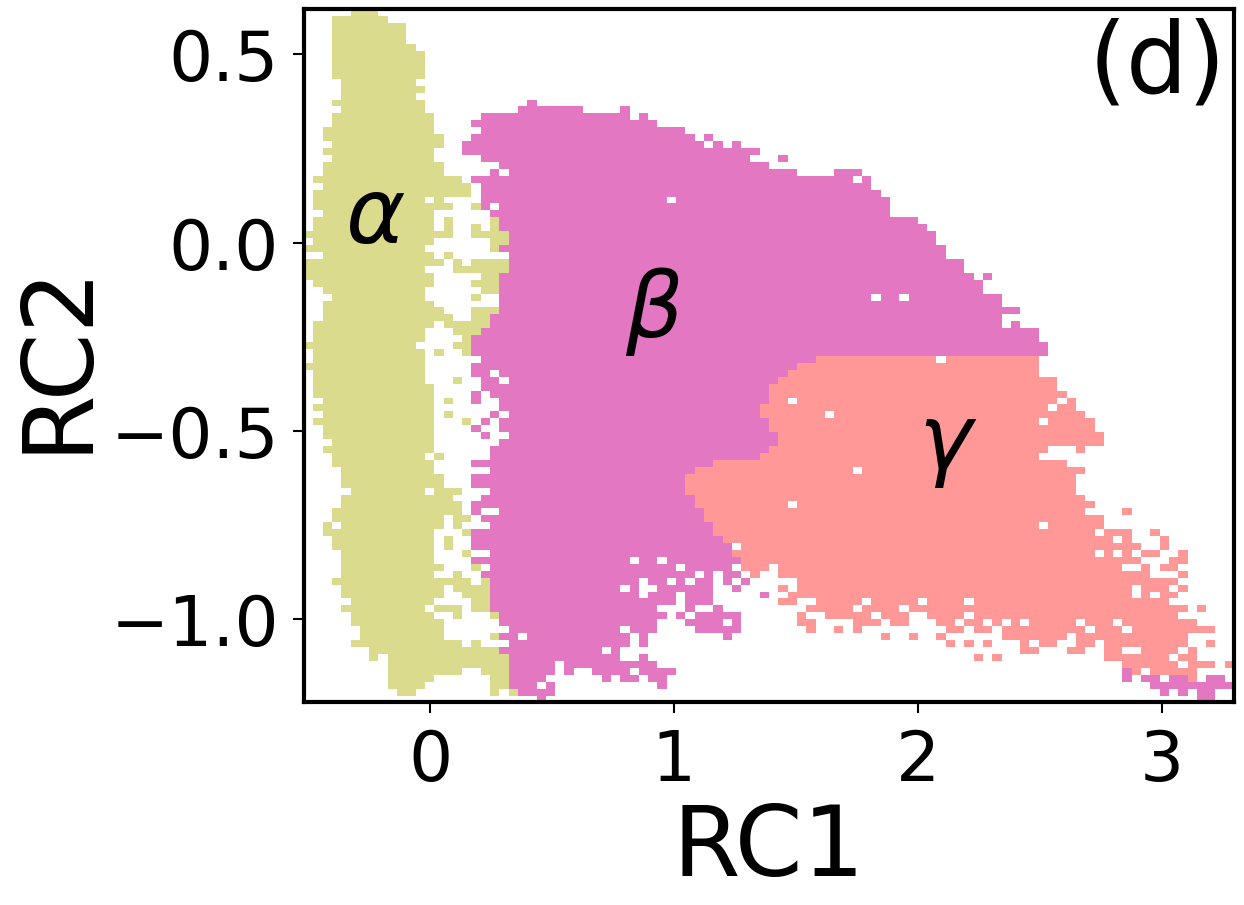}
    \end{minipage}
    \caption{\textbf{Free energies and SPIB analysis.} Reweighted free energy from WTmetaD simulations performed with the SS model (a) along $\bar{\theta}_1$ and (b) in the $(\bar{\theta}_1,N_{8+})$ space. The plotted states are limited to those involved in the transitions between the liquid-like state and form I. (c) The state labels learnt by SPIB projected into the $(\bar{\theta}_1,N_{8+})$ space. (d) The state labels in the two-dimensional space of the SPIB-learnt RCs. Here the three colors mustard, fuchsia and orange denote three different metastable states detected by SPIB.}
    \label{fig:SPIB}
\end{figure*}

\begin{figure*}[ht]
    \centering
    \begin{minipage}[b]{0.35\textwidth}
    \includegraphics[width=\textwidth]{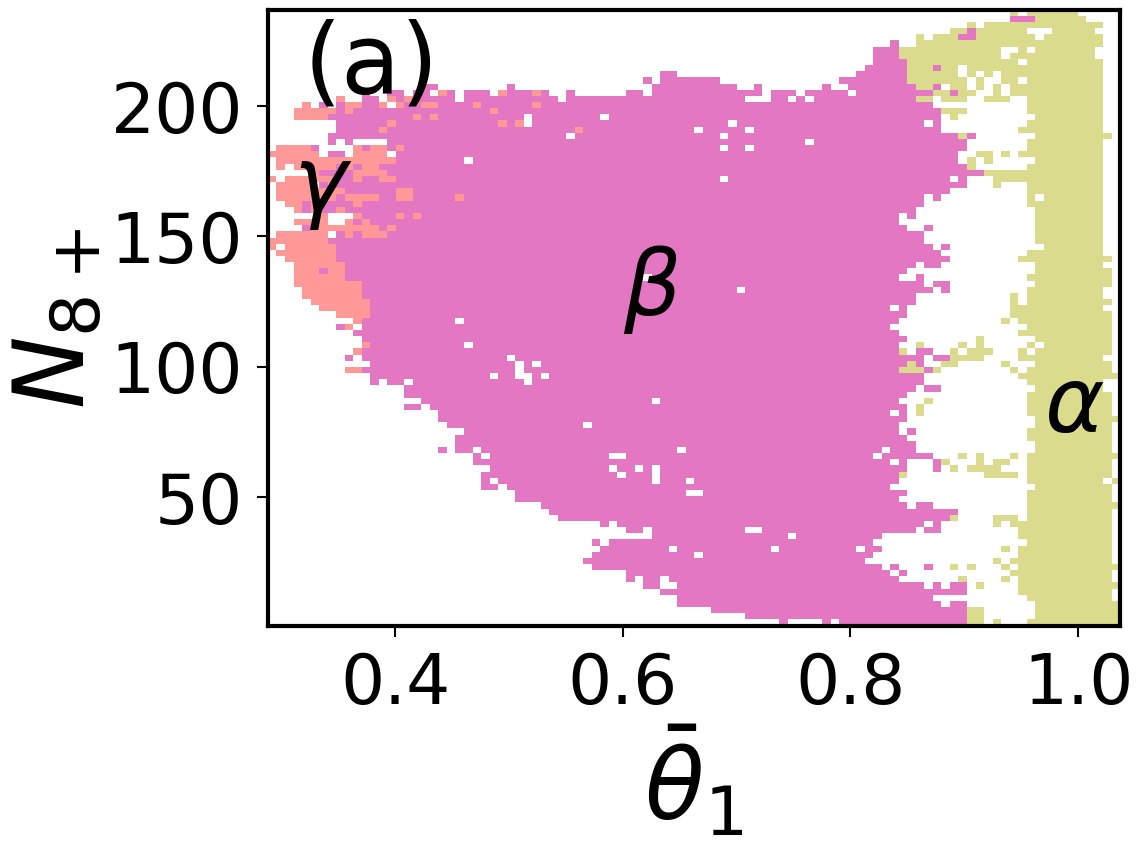}
    \end{minipage}
    \begin{minipage}[b]{0.35\textwidth}
    \includegraphics[width=\textwidth]{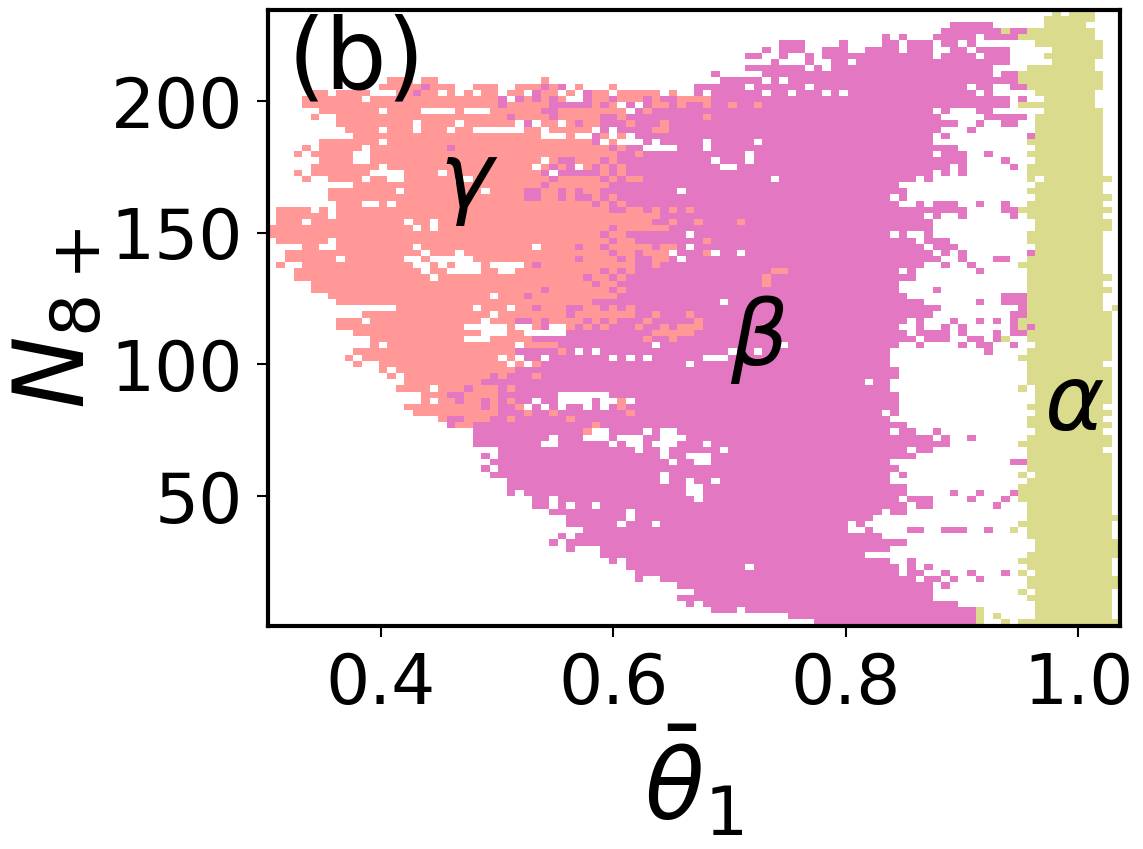}
    \end{minipage}
    \caption{\textbf{Metastable states comparison.} The state labels learnt by SPIB in the $(\bar{\theta}_1,N_{8+})$ space for (a) the GT model and (b) the full model. The three colors mustard, fuchsia and orange denote three different metastable states detected by SPIB.}
    \label{fig:labels}
\end{figure*}

	\end{document}

\title{\LARGE Supplementary Information}

 \author{Renjie Zhao}
 \affiliation{Chemical Physics Program and Institute for Physical Science and Technology, University of Maryland, College Park 20742, USA.}

\author{Ziyue Zou}
 \affiliation{Department of Chemistry and Biochemistry, University of Maryland, College Park 20742, USA.}

 \author{John D. Weeks*}
 \email{jdw@umd.edu}

\affiliation{Institute for Physical Science and Technology and Department of Chemistry and Biochemistry, University of Maryland, College Park 20742, USA.}

 \author{Pratyush Tiwary*}
 
 \email{ptiwary@umd.edu}
 \affiliation{Institute for Physical Science and Technology and Department of Chemistry and Biochemistry, University of Maryland, College Park 20742, USA.}

\maketitle

\section*{Supplementary Notes}

\subsection*{Note 1: Averaged intermolecular angles}
For an arbitrary solute molecule $i$, we have
\begin{equation}
\hat{\theta}(i)=\frac{\sum_{j}\sigma(r_{ij})\frac{1}{2}[(\pi-2\theta) \tanh\funcapply(5\theta-9.25)+\pi]}{\sum_{j}\sigma(r_{ij})},
\label{eq:angle}
\end{equation}
where $\theta$ is the intermolecular angle between characteristic vectors on molecule $i$ and molecule $j$, and $\sigma(r_{ij})$ is a switching function of intermolecular distance $r_{ij}$. In order to eliminate the mirror image symmetry, the hyperbolic tangent switching function is implemented in Eq.~\ref{eq:angle}. We then compute the mean of $\hat{\theta}(i)$ over the group of solutes to obtain the averaged intermolecular angles $\bar{\theta}_1$ and $\bar{\theta}_2$, for which the subscripts refer to the specific intermolecular angles defined in the context of pair orientational entropy. $\mu^2_{\theta_1}$ and $\mu^2_{\theta_2}$ are computed as the second moments of the $\hat{\theta}_1(i)$ and $\hat{\theta}_2(i)$ distributions.

\subsection*{Note 2: Coordination number}
We calculate the coordination number of urea molecules by the following continuous and differentiable expression,
\begin{equation}
c(i)=\sum_{j}\frac{1-(r_{ij}/r_\text{c})^6}{1-(r_{ij}/r_\text{c})^{12}},
\label{eq:CN}
\end{equation}
where $r_{ij}$ is the distance between reference sites of molecules $i$ and $j$, and $r_\text{c}$ is the cutoff. $N_{8+}$, $N_{11+}$, which are the populations of molecules with coordination numbers greater than 8 and 11, and the second moment of coordination numbers $\mu^2_c$ are derived from the $c(i)$ distribution.

\subsection*{Note 3: Distributions of $N_{8+}$}
\begin{figure}[ht]
    \centering
    \begin{minipage}[b]{0.5\textwidth}
        \begin{picture}(0,0)
            \put(-135,0){\large(a)} % Adjust position as needed
        \end{picture}
        \includegraphics[width=\textwidth]{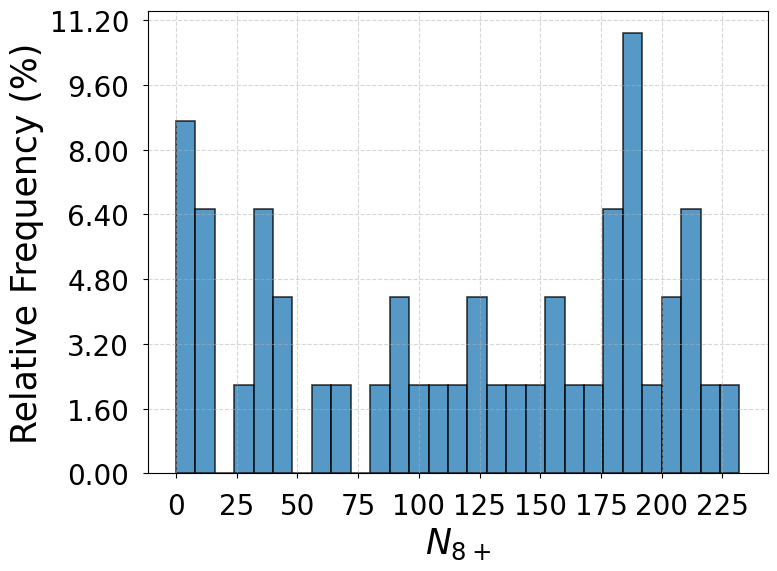}
    \end{minipage}
    \begin{minipage}[b]{0.5\textwidth}
        \begin{picture}(0,0)
            \put(-135,0){\large(b)} % Adjust position as needed
        \end{picture}
        \includegraphics[width=\textwidth]{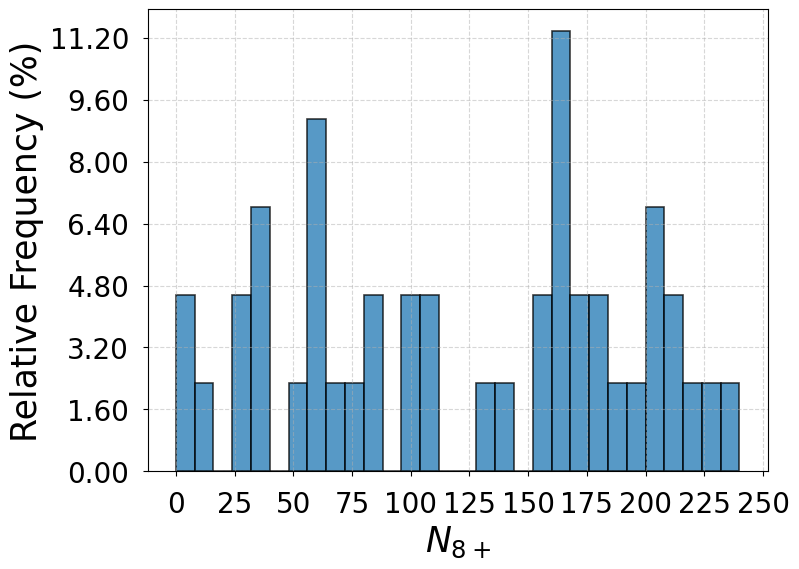}
    \end{minipage}
    \begin{minipage}[b]{0.5\textwidth}
        \begin{picture}(0,0)
            \put(-135,0){\large(c)} % Adjust position as needed
        \end{picture}
        \includegraphics[width=\textwidth]{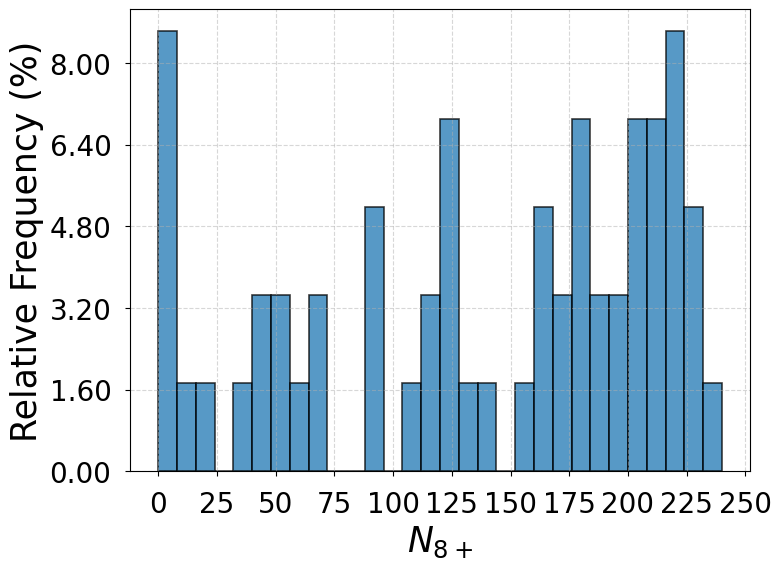}
    \end{minipage}
    \caption{\textbf{Cluster size information during transitions.} Distributions of $N_{8+}$ occurring with transitions between the liquid-like basin and the primary crystal basin from WTmetaD simulations performed with (a) the SS model, (b) the GT model, and (c) the full model.}
    \label{fig:N8}
\end{figure}
In Supplementary Fig. 1 are the distributions of $N_{8+}$ when the system transitions between the liquid-like basin and the primary crystal basin for the SS model, the GT model, and the full model. Based on the criterion that the liquid-like configurations from WTmetaD simulations count toward the liquid state when their $N_{8+}$ values are within the three-sigma limits of the liquid state $N_{8+}$ distribution drawn from a 50 ns unbiased simulation and otherwise count toward the dense liquid state, the transitions between the liquid-like basin and the primary crystal basin take place directly from or to the liquid state in $8.7\%$ of the total 46 events for the SS model, $4.5\%$ of the total 44 events for the GT model, and $8.6\%$ of the total 58 events for the full model. The other transitions are mediated by the dense liquid state. The results from the three models consistently indicate a predominant two-step nucleation mechanism.

The distributions of $N_{8+}$ beyond the dense liquid state threshold ($\sim9.61$) do not necessarily take similar shapes for the three models. This is because there can exist one or multiple clusters in different configurations and $N_{8+}$ is not accurately proportional to the size of the cluster.

\section*{Supplementary Figure}